\documentclass[aps,prc,twocolumn,groupedaddress,showpacs]{revtex4}

\usepackage{graphicx}
\usepackage{dcolumn}
\usepackage{bm}

\def\F{{\cal F}}

\begin{document}

\preprint{Cyclotron Institute at Texas A\&M University}

\title{Precise Half-Life Measurement of the \\
	Superallowed $\beta^+$ emitter $^{10}$C}

\author{V.E. Iacob}
\email[Email address:~]{iacob@comp.tamu.edu}
\author{J.C. Hardy}
\author{V. Golovko}
\author{J. Goodwin}
\author{N. Nica}
\author{H.I. Park}
\author{L. Trache}
\author{R.E. Tribble}

\affiliation{Cyclotron Institute at Texas A\&M University}

\date{\today}

\begin{abstract}
The half-life of $^{10}$C has been measured to be 19.310(4)~s, a result with 0.02\%
precision, which is a factor of three improvement over the best previous result.  Since
$^{10}$C is the lightest superallowed $0^+ \rightarrow 0^+$ $\beta^+$ emitter, its $ft$ value
has the greatest weight in setting an upper limit on the possible presence of scalar
currents.

\end{abstract}


\maketitle

\section{Introduction \label{intro}}
For many years, measurements of superallowed $0^+ \rightarrow 0^+$ $\beta^+$ transitions
have resulted in increasingly precise tests of several key ingredients of the
electroweak standard model: the Conserved Vector Current (CVC) hypothesis, the absence
of scalar currents and the unitarity of the Cabibbo-Kobayashi-Maskawa (CKM) matrix.  In
particular, a recent survey \cite{HT05} of world data on superallowed decays confirmed
CKM unitarity (to $\pm 0.1$\%) and two of the CVC assertions: that the vector coupling
constant has the same value for all nine well-known transitions ($\pm 0.01$\%), and that
the induced scalar interaction is consistent with zero ($\pm 0.1$\%).  Furthermore, the
latter result also ruled out any fundamental scalar interaction at the same level of
precision.

All these are important results, but it is the test for scalar currents that motivates
the present measurement.  Each superallowed transition is characterized by an $\F t$
value, which is derived from its measured $ft$ value by the inclusion of calculated
corrections for radiative and isospin-symmetry-breaking effects \cite{HT05}. The test
for scalar currents is based on how the $\F t$ values vary with the $Z$ of the daughter
nucleus: if there is no scalar current, then $\F t$ should be completely independent of
$Z$; if a scalar current were to exist, then the locus of $\F t$ values would show a slope
(either up or down) that increases strongly -- and would be most easily identifiable -- at low
$Z$ \cite{HT05a}.  The decay of $^{10}$C is the lowest-$Z$ example of a superallowed
$0^+ \rightarrow 0^+$ $\beta^+$ transition and thus is the most influential in probing for
a scalar current.

The $ft$ value that characterizes any $\beta$-transition depends on three measured
quantities: the total transition energy, $Q_{EC}$, the half-life, $t_{1/2}$, of
the parent state and the branching ratio, $R$, for the particular transition of
interest.  The $Q_{EC}$-value is required to determine the statistical rate function,
$f$, while the half-life and branching ratio combine to yield the partial half-life,
$t$.  The current $Q_{EC}$ value for $^{10}$C yields a value for $f$ with 0.05\%
precision; the branching ratio is known to 0.13\%; and, before the
measurement reported here, the half life was known to 0.06\% \cite{HT05}.  All these values
can be improved.  Our new half-life value for $^{10}$C has 0.02\% precision and
represents a first step in bringing the uncertainty in the $ft$ value for this
transition down from 0.15\%, where it is now, to 0.04\% or better, where the most
precisely known $ft$ values now lie.  Such increased precision would correspondingly
increase the sensitivity of the corrected $\F t$ value to the presence of a scalar
interaction.

\section{Execution of the experiment }

Our experimental set-up was similar to the one used in our recently reported measurement
of the half-life of $^{34}$Ar \cite{Ia06}.  Apart from the details specific to the
$^{10}$C measurement, only a brief description will be presented here.

A primary beam of $^{11}$B at 23 $A$MeV, produced by the superconducting cyclotron
at Texas A\&M University, impinged on a 1.6-atm hydrogen gas target cooled to liquid
nitrogen temperature.  The fully stripped ejectiles were then analyzed by the
Momentum Achromat Recoil Separator (MARS), with the result that a $>$99.8\% pure beam of
radioactive $^{10}$C at 18.5 $A$MeV was produced at the focal plane.  This beam
exited into air through a 51-$\mu$m-thick kapton window, and then passed through a
0.3-mm thin BC-404 plastic scintillator and through a set of aluminum degraders, 
eventually being implanted in the 76-$\mu$m-thick aluminized mylar tape of a fast
tape-transport system.  The combination of $m/q$ selectivity in MARS and range
selectivity in the degraders resulted in no detectable impurities being present in the
implanted samples.

\begin{figure}[t]
 \includegraphics[width=\columnwidth]{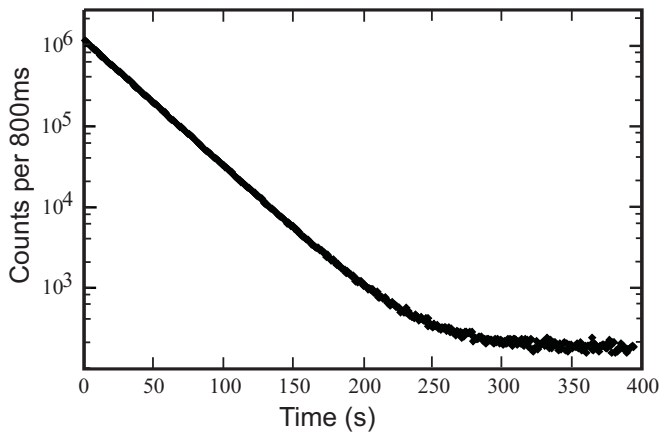}%
 \caption{\label{fig1} Total time decay spectrum of $^{10}$C.}
\end{figure}

The $^{10}$C nuclei were collected in the tape for 10 s, then the beam was turned off
and the activity was moved in 180\,ms to the center of a 4$\pi$ proportional gas counter.  
The gas counter was shielded against neutrons and gammas to reduce the background rate
to a minimum.  Once the activity arrived at the gas counter, the emitted positrons were
detected with $>$95\% efficiency and their signals multiscaled for 400\,s, which is just
over 20 half-lives of the 19.3-s $^{10}$C.  The clock-controlled collect-move-detect
cycles were repeated until the desired statistics were achieved.  The time base for the
acquired spectra was defined by a Stanford Research System pulser (Synthesized Function
Generator, model DS345), which is accurate to 0.01 ppm.  For this measurement, we recorded
in this way a total of $8.5 \times 10^7$ decays.  The total decay spectrum, spanning nearly
four decades in count rate, is presented in Figure\,\ref{fig1}.

In order to probe as carefully as possible for weak long-lived impurities, we also made
measurements with two different collect-move-detect settings of 20\,s-0.180\,s-900\,s and
120\,s-0.180\,s-900\,s.  Finally, we established room background under measurement
conditions by recording data with the ``normal" collect-move-detect settings but with
the tape-move feature disabled. 

Overall, great care was taken to test thoroughly for systematic effects caused by detection
parameters.  Signals from the gas detector were amplified and sent to a discriminator,
whose output was split and processed in parallel by two separate systems, each with a different
pre-set dominant non-extendable dead-time.  The two data streams allowed us to test that our
dead-time-corrected result was independent of the actual dead time of the circuit.  In
addition to this check, the overall measurement was also subdivided into many separate
runs, differing from one another only in their detection parameters: dominant dead-time, 
detector bias and discriminator threshold.

Other special precautions were taken too.  The tape-transport system is quite consistent
in placing the collected source within $\pm$3 mm of the center of the detector, but it is
a mechanical device, and occasionally larger deviations occur.  Although there is no
reason to expect that this could lead to a false half-life result, we separately recorded
the number of implanted nuclei detected in the scintillator at the beginning of each cycle, 
and the number of positrons recorded in the gas counter during the subsequent count period.  
The ratio of the latter to the former is a sensitive measure of whether the source is
seriously misplaced in the proportional counter.  In analyzing the data later, we rejected
the results from any cycle with an anomalous (low) ratio.  This happened rarely: less than
1\% of the cycles were rejected for this reason.

\section{Results}

We analyzed the two measurements made with longer collect and detect times first in order to search
for possible long-lived impurities.  Only one very weak contaminent activity could be detected,
and it proved not to have originated from the collected sample but instead from neutron
activation of the copper housing of the gas detector.  We identified 5.12-min $^{66}$Cu, produced
from $^{65}$Cu(n,$\gamma$)$^{66}$Cu.  Although $^{64}$Cu was undoubtedly also produced from neutron
capture on the other stable isotope of copper, $^{63}$Cu, only $^{66}$Cu needed to be considered
in our subsequent analyses since $^{64}$Cu has a half-life of 12.7~hours and thus serves only to
increase slightly the constant background.  We determined that, under the timing conditions of our
$^{10}$C half-life measurements, the activity of $^{66}$Cu produced by neutron activation was
0.03\% of the activity of the $^{10}$C sample implanted in the tape each cycle.  Although very
small, this $^{66}$Cu contribution was nevertheless included in all our $^{10}$C decay analyses.

\begin{figure}[b]
 \includegraphics[width=\columnwidth]{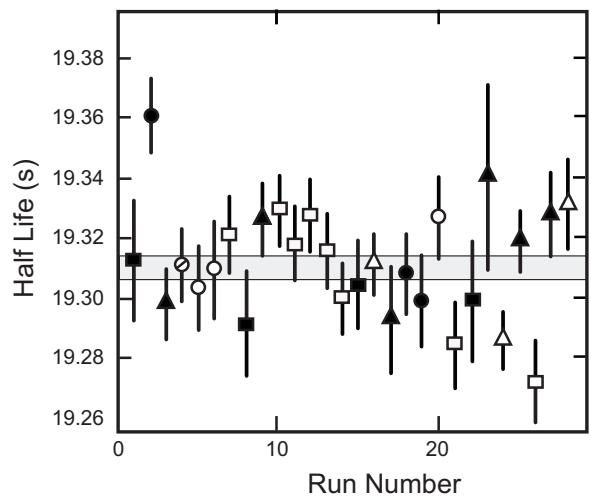}%
 \caption{\label{fig2} Test for possible systematic bias in the $^{10}$C half-life  
 measurement due to discriminator threshold or detector bias. The circles, squares and
 triangles represent discriminator settings of 150, 200 and 250 mV, respectively.  The
 full symbols signify a detector bias of 2500 or 2550 V; the open ones signify 2600 or
 2650 V; the open symbol with a diagonal line through it represents a bias of 2400 V.  
 The shaded band gives one standard deviation around the average half-life.  The quality
 of the fit to a single average value is given by $\chi^2$/ndf = 38.6/27.}
\end{figure}

In all, we recorded data from 478 collect-move-detect cycles, divided into 28 separate runs, each
characterized by a different combination of detector high-voltage, discriminator-threshold and dominant
dead-time settings.  We processed the data run-by-run, first by summing the dead-time-corrected
spectra from all accepted cycles in each run.  Then the 28 sum spectra were individually analyzed
by a maximum-likelihood fit with a function incorporating two exponentials -- for $^{10}$C and
$^{66}$Cu -- and a constant background.  (The methods used were those described in
Ref.\,\cite{Ko97}.)  The $^{66}$Cu parameters in each run were fixed to
its accepted half-life, 5.120(14) min \cite{Bh98}, and to an intensity determined by adjustment
of the results from our contaminant tests to account for the actual collect and detect times of
the particular run.

The half-life results from all 28 runs are plotted individually in Figure\,\ref{fig2}, where
each point is identified with two of its parameter settings.  It can be seen that there is no
indication of a systematic dependence on detector bias or discriminator threshold.  The
results, which had already been dead-time corrected, were equally insensitive to the
actual circuit dead time used in each run, although this parameter is not identified in
the figure.

\begin{figure}[t]
  \includegraphics[width=\columnwidth]{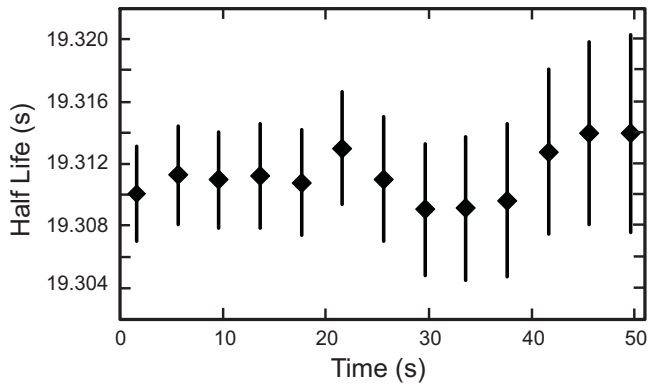}%
 \caption{\label{fig3} Test for possible systematic errors in the
	measurement of the decay of $^{10}$C caused by undetected 
	short-lived impurities or by rate-dependent counting
	losses. The abscissa represents the time period at the
	beginning of the counting cycle for which the data were
	omitted from the fit.
	}
\end{figure} 

\begin{table}[b]
\caption{Error budget for the $^{10}$C half-life measurement
\label{table1}}
\vskip 1mm
\begin{ruledtabular}
\begin{tabular}{lc}
Source   & Uncertainty (ms) \\
\hline \\[-2.5mm]
counting statistics		&  2.6	\\
$^{66}$Cu contaminant	&  1.5	\\
dead time		&  0.6	\\
detector bias   	&  2.0	\\
discriminator threshold &  2.5	\\[2mm]
total uncertainty (ms)	& 4.4 \\ 
\end{tabular}
\end{ruledtabular}
\end{table} 

As a further systematic check, in this case on the fitting procedure, we generated by Monte Carlo
techniques a set of spectra which matched, cycle by cycle, the statistics and composition of the
data from each cycle.  The half-life we used in generating these spectra was chosen to be similar
to that of $^{10}$C.  We then fitted these artificial data with the same techniques we used for
the real data.  The half-life we obtained from fitting our artificial data agreed with the input
half-life within statistical uncertainties, thus validating the fitting procedure and the result
obtained from it.

We also checked for the possible presence of unidentified short-lived impurities or other possible
count-rate-dependent effects.  We removed data from the first 4$\,$s of the counting period in
each measurement and refitted the remainder; then we repeated the procedure, removing the first
8$\,$s, then 12$\,$s and so on.  As can be seen from Fig. \ref{fig3}, within statistics the derived
half-life was also stable against these changes. 

Our final result for the half life of $^{10}$C is 19.310(4)\,s.  The quoted uncertainty contains
contributions from a number of sources as itemized in Table~\ref{table1}.  We determined the
contribution from each experimental parameter -- dead time, detector bias or discriminator
threshold -- by the spread in half-life averages obtained when all the data were grouped only
according to the values set for that parameter.  We obtained the contribution from the $^{66}$Cu
contaminant by assuming that the uncertainty of its intensity equalled one-third of its
total value.  In all cases, we consider these to be very conservative estimates of the experimental
and systematic uncertainties.

\section{Conclusions}

We have obtained the most precise value yet for the $^{10}$C half life, 19.310(4)\,s.  This
result -- with 0.02\% precision -- can
be compared with the only two previous measurements whose precision is within a factor of ten
of that of the present measurement \cite{HT05}: 19.280(20)\,s \cite{Az74} and
19.295(15)\,\cite{Ba90}.  Our value agrees with the more
recent, 1990, result \cite{Ba90} within the latter's uncertainty, and is not seriously different
from the 1974 measurement \cite{Az74}.  However, our precision is better than both by at least
a factor of four.  When all three results are averaged together, the resulting half-life is
19.308(4)\,s and, with the other properties for the superallowed decay of $^{10}$C taken from
Ref.\,\cite{HT05}, the corresponding $ft$ value becomes 3042.4(43)\,s.  This represents a reduction in
uncertainty from the previous $ft$ value \cite{HT05}, 3039.5(47)\,s, now leaving the branching
ratio and, to a lesser extent, the $Q_{EC}$ value as the major contributors to the overall
uncertainty.  

It is interesting to note that our new measurement of the half life increases the $ft$ value
for $^{10}$C as well as reducing its uncertainty.  Naturally, its corrected $\F t$ value is
increased as well, to 3077.4(46)\,s.  This is slightly above the overall average of all well-known
superallowed transitions \cite{HT05} and, if this tendency for $^{10}$C
is reenforced by branching-ratio and $Q_{EC}$-value measurements with improved precision,
it could indicate the presence of a small contribution from a scalar current.  Clearly, high
priority should be attached to the remeasurement of these two quantities with improved precision.

\acknowledgements
This work was supported by the U.S. Department of Energy under Grant No.~DE-FG03-93ER40773 and
by the Robert A. Welch Foundation under Grant No. \mbox{A-1397}.

\end{document}